\documentclass[twocolumn,pra,superscriptaddress]{revtex4}
\usepackage{amsmath,amssymb,}
\usepackage{mathrsfs}
\usepackage{graphicx}% Include figure files
\usepackage{dcolumn}% Align table columns on decimal point
\usepackage{bm}% bold math
\usepackage[colorlinks,citecolor=red]{hyperref}

\usepackage{rotating}
\usepackage{floatrow}

\usepackage{lipsum}

\usepackage{bbm}

\begin{document}

\title{Damping transition in an open generalized Aubry-Andr\'e-Harper model}

\author{Peng He}
\author{Yu-Guo Liu}
\author{Jian-Te Wang}
\affiliation{School of Physics, Nanjing University, Nanjing 210093, China}
\affiliation{National Laboratory of Solid State Microstructures, Collaborative Innovation Center of Advanced
Microstructures, Nanjing University, Nanjing 210093, China}

\author{Shi-Liang Zhu}
	\email{slzhu@nju.edu.cn}
\affiliation{Guangdong Provincial Key Laboratory of Quantum Engineering and Quantum Materials, School of Physics and Telecommunication Engineering, South China Normal University, Guangzhou 510006, China}

\affiliation{Guangdong-Hong Kong Joint Laboratory of Quantum Matter, Frontier Research Institute for Physics, South China Normal University, Guangzhou 510006,
China}

\date{\today}

\begin{abstract}
We study the damping dynamics of the single-particle correlation for an open system under periodic and aperiodic order, which is dominated by the Lindblad master equation. In the absence of the aperiodic order, the Liouvillian superoperator exhibits the non-Hermitian skin effect, which leads to unidirectional damping dynamics, dubbed as ``chiral damping". Due to the non-Hermitian skin effect, the damping dynamics is boundary sensitive: The long-time damping of such open systems is algebraic under periodic boundary conditions but exponential under open boundary conditions. We reveal the phase transition with the inclusion of the hopping amplitude modulation. By using the spectral topology and a finite-size scaling analysis in the commensurate case, we  show there exists a phase transition of the skin effect with non-Bloch anti-parity-time symmetry breaking. For the incommensurate case, we find richer phases with the coexistence of the non-Hermitian skin effect and the Anderson localization, which are separated by a generalized mobility edge. We reveal the transition of the damping dynamics as a consequence of the phase transition. Furthermore, we propose a possible scheme with ultracold atoms in a dissipative momentum lattice to realize and detect the damping dynamics.

\end{abstract}
%\pacs{42.50.Pq, 37.30.+i, 03.67.Bg, 76.30.Mi}

\maketitle

\section{Introduction}
With advances in manipulating dissipation and quantum coherence in the laboratory, the past years have seen
a revived interest in the theory of open and non-equilibrium systems \cite{Bergholtz2021,Ashida2020}. Effective non-Hermitian descriptions have prominently transpired in a plethora of non-conserved systems, such as classical waves with gain and loss \cite{XZhu2014,Popa2014,HZhou2018,LFeng2014,Regensburger2012,Cerjan2019}, solids with finite quasiparticle lifetimes \cite{Kozii2017,Michishita2020}, and open quantum systems with Markovian reservoirs \cite{LXiao2020,LXiao2021} \emph{etc}. Unique features of non-Hermitian systems have been recognized in various physical contexts, especially in the topological bands \cite{HShen2018,SYao2018,ZGong2018,Kawabata2019}. For instance, non-Hermitian systems generally possess complex-valued spectra, classified by the homotopy group of the general linear group GL(n,$\mathbb{C}$) \cite{Kawabataprl,ZGong2018}. The non-Hermitian systems can be gapped in two distinct ways, with a line gap or a point gap. A hallmark of the point-gap topology is failure of the Bloch theorem and the non-Hermitian skin effect (NHSE), namely, the anomalous boundary localization for majority of bulk states \cite{SYao2018,Kunst2018,KZhang2020,Borgnia2020,Okuma2020,Yokomizo2019,CHLee2019}.

On the other hand, localization has long time been recognized as important physical implication of scattering, transmission, or interference of waves in dissipative media, since the discovery of Anderson localization \cite{Anderson1958,Abrahams1979}. The Anderson localization can occur for lattice with disorder and long-range aperiodic order \cite{Anderson1958,Abrahams1979,Harper1955,Aubry1980,Hofstadter1976,Jitomirskaya1999}. In recent years, there is growing attention on the interplay of non-Hermitian physics and disorder effect \cite{Longhi2019,Longhi2021,DWZhang2020,QBZengprr,QBZengprb,HJiang2019,CZhang2021,YLiu2020,
DWZhang2020SChina,Kawabata2021,GQZhang2021,LZTang2021,LFZhang2021,LZTang2020}. One line of work is the characterization and classification of the matter phase in terms of disorder \cite{Kawabata2019,Lugwig2016,XLuo2021}. Other topics concern cooperation with coherent control techniques \cite{LZhou2021}. Disorder or quasiperiodicity leads to exotic behaviors, including localization-delocalization transition under the parity-time ($\mathcal{PT}$) symmetry breaking \cite{Hatano1996,Hatano1998,Hamazaki2019}, generalized mobility edges \cite{QBZengprr} and anomalous particle transport \cite{Longhi2021}, among which the non-Hermitian Aubry-Andr\'e-Harper (AAH) model provides as a paradigmatic example. However, yet most works only concentrate on non-Hermitian Hamiltonian problems, systems resting on Liouvillians are still rarely studied.

In this paper, we consider an open quantum system with periodic and aperiodic orders, governed by the Lindblad master equation. The aperiodic order is introduced by the modulation of lattice hopping amplitude. Following the methods developed in Refs. \cite{FSong2019,CHLiu2020}, we study the dynamics of this system in terms of the damping matrix derived from the Liouvillian. The damping matrix is mathematically non-Hermitian, and can be seen as a generalized AAH model. For the commensurate case, we show a phase transition of the skin effect with non-Bloch anti-$\mathcal{PT}$ symmetry breaking (reminiscent of the concept of non-Bloch $\mathcal{PT}$ symmetry, see Refs. \cite{Longhioe,Longhiprr}). The phase transition is characterized by both the spectral topology and mean inverse participation ratio, further revealed by a finite-size scaling analysis. As a physical consequence, we find transition of the damping dynamics across distinct phases. To characterize the damping transition, we investigate the longest relaxation time and relaxation velocity, showing a way to control the relaxation process in the experiment. As revealed in Ref. \cite{FSong2019}, in the absence of the modulation, the system will undergo chiral damping, i.e., the system starts to damp in an unidirectional way. With the increase in modulation strength, the chiral damping fades away. For the incommensurate case, we find richer phases due to the competition of the non-Hermitian skin effect and the Anderson localization. We identify phases with the coexistence of two types of the states, which are separated by a generalized mobility edge. Furthermore, we propose a possible scheme based on the momentum lattice to realize and detect our model \cite{Gadway2015,Meier2016,Meiernc}.

The rest of this paper is organized as follows. In Sec. \ref{sec2}, we briefly review the general framework on how to cast a Liouvillian with linear jump operators to a non-Hermitian damping matrix. In Sec. \ref{sec3}, we contract a generalized AAH model, based on a one-dimensional dimerized lattice with modulation of intra-cell hopping amplitude, then reveal the phase transition in terms of the properties of the spectra and the eigenmodes.In Sec. \ref{sec4}, we numerically calculate the time evolution of our model, and study the dynamical phase transition.  In Sec. \ref{sec5}, we consider the incommensurate case, and identify three distinct phases and the existence of a generalized mobility edge. Then we present an experimental proposal for realizing and detecting this model in Sec. \ref{sec6}. Finally, a short summary is given in Sec. \ref{sec7}.

\section{General formalism}\label{sec2}
We start with outlining the Lindblad damping matrix framework for open quantum systems,  following Ref. \cite{FSong2019}. The dynamics of an open system undergoing Markovian damping is governed by the following Lindblad master equation,
\begin{equation}
\begin{aligned}
\frac{d\rho}{dt} &=-i[H, \rho]- \sum_{\mu}(L_{\mu}^{\dagger} L_{\mu} \rho+\rho L_{\mu}^{\dagger} L_{\mu}-2 L_{\mu} \rho L_{\mu}^{\dagger}) \\
&=-i (H_{\mathrm{eff}} \rho-\rho H_{\mathrm{eff}}^{\dagger} )+2 \sum_{\mu} L_{\mu} \rho L_{\mu}^{\dagger}\,,
\end{aligned}\label{eq_master}
\end{equation}
where $\rho$ is the density matrix, $H$ is the Hamiltonian, and $L_\mu's$ are the Lindblad jump operators describing coupling to environment. For short-time dynamics before any quantum jump event, when the last term in Eq. (\ref{eq_master}) is negligible, the system is described by an effective non-Hermitian Hamiltonian $H_{\mathrm{eff}}=H-i\sum_{\mu} L_\mu^\dagger L_\mu$.

We consider noninteracting particles in a tight-binding lattice. The Hamiltonian can be generally written as $H=\sum_{ij} h_{ij} c_i^\dagger c_j$, where $c_{i}^\dagger~(c_i)$ is the creation (annihilation) operator on lattice site $i$, and $h_{ij}=h_{ji}^*$  is a
time-dependent hopping amplitude $(i \neq j)$ or onsite potential
$(i = j)$. To see the full-time evolution of the density matrix, it is convenient to define the single-particle correlation $\Delta_{ij}=\mathrm{Tr}[c_i^\dagger c_j\rho(t)]$. The time evolution then follows $d\Delta_{ij}/dt=\mathrm{Tr}[c_i^\dagger c_j d\rho/dt]$.
If only the single-particle gain and loss with linear gain dissipator $L_\mu^g=\sum_i D_{\mu i}^g c_i^\dagger$ and loss dissipator $L_\mu^l=\sum_i D_{\mu i}^l c_i$ are concerned, the evolution equation can be recast as,
\begin{equation}
\frac{d \Delta(t)}{d t}=X \Delta(t)+\Delta(t) X^{\dagger}+2 M_{g}\,,\label{eq_evo}
\end{equation}
where $X\equiv i h^{T}-(M_{l}^{T}+M_{g})$ is dubbed the damping matrix with $(M_g)_{ij}\equiv \sum_\mu D_{\mu i}^{g*}D_{\mu j}^g$ and $(M_l)_{ij}\equiv \sum_\mu D_{\mu i}^{l*}D_{\mu j}^l$. Deduction from the single-particle correlation of its steady value $\tilde{\Delta}(t)=\Delta(t)-\Delta_s$, homogenizing Eq. (\ref{eq_evo}) gives
\begin{equation}
\tilde{\Delta}(t)=e^{X t} \tilde{\Delta}(0) e^{X^{\dagger} t}.\label{eq_delta}
\end{equation}
Here steady state correlation $\Delta_s=\Delta(\infty)$ is determined by $d\Delta_s/dt=0$, or $X \Delta_{s}+\Delta_{s} X^{\dagger}+2 M_{g}=0$.

\section{Model}\label{sec3}
We consider the following dimerized AAH model \cite{Ganeshan2013},
\begin{equation}
H=\sum_{i=1}^{N}(t_{1}+\lambda_i ) \hat{c}_{i, A}^{\dagger} \hat{c}_{i, B}+t_{2} \hat{c}_{i, B}^{\dagger} \hat{c}_{i+1, A}+\mathrm{h.c.}\,,\label{eq_ham}
\end{equation}
where $A$ and $B$ denote two internal degrees of freedom, $\lambda_i=\lambda \cos(2\pi\alpha i+\delta)$ depicts modulation of intracell hopping with real parameters $\lambda$, $\alpha$, and $\delta$. When $\alpha=p/q$  (with $p$ and $q$ being relatively prime positive integers), the lattice has an enlarged periodicity over $q$ cells, whereas the lattice becomes quasiperiodic with the incommensurate modulation when $\alpha$ is an irrational number. Each unit cell contains a single loss and gain dissipator,
\begin{equation}
L_{x}^{l}=\sqrt{\gamma_{l} / 2} (c_{x A}-i c_{x B} )\,,~~L_{x}^{g}=\sqrt{\gamma_{g} / 2} (c_{x A}^{\dagger}+i c_{x B}^{\dagger} )\,,\label{eq_jump}
\end{equation}
where $x$ denotes the lattice site. For simplicity, we firstly study the commensurate case. In the commensurate case, the damping matrix is translational invariant with respect to $q$ cells under the periodic boundary condition (PBC). In the basis,
\begin{equation}
\hat{c}_{k}=(\hat{c}_{1 k}~~e^{-i k / q}\hat{c}_{2 k}~~ \cdots ~~e^{-i(q-1) k / q} \hat{c}_{q k})^{\mathrm{T}}\,,
\end{equation}
the damping matrix in momentum space can be written as
\begin{equation}
\begin{split}
X(k)_{mn}=&i(\delta_{m,n-1}\overline{t}_m+\delta_{m-1,n}\overline{t}_n'+\delta_{m,1}\delta_{n,2q}t_2 e^{-ik}\\&+\delta_{m,2q}\delta_{n,1}t_2 e^{ik})-\frac{\gamma}{2} \mathbb{I}_{2q}\,,\label{eq_xk}
\end{split}
\end{equation}
where $k\in [0,2\pi]$, $\gamma=\gamma_l+\gamma_g$, $\mathbb{I}_{2q}$ is a $2q\times 2q$ identity matrix, $\overline{t}_{2i-1}=t_1+\gamma/2+\lambda_{2i-1}$, $\overline{t}_{2i-1}'=t_1-\gamma/2+\lambda_{2i-1}$, and $\overline{t}_{2i}(t_{2i}')=t_2$. Here to shorten notations we define $\hat{c}_{1 k}=(\hat{c}_{1A k} ~\hat{c}_{1B k})$.

We can write $X$ in terms of its right and left eigenvectors,
\begin{equation}
X=\sum_{n} \lambda_{n}|u_{R n}\rangle\langle u_{L n}|\,,
\end{equation}
where $X^\dagger|u_{L n}\rangle=\lambda_n^*|u_{L n}\rangle$ and $X|u_{R n}\rangle=\lambda_n |u_{R n}\rangle$. Then we can also reexpress Eq. (\ref{eq_delta}) as
\begin{equation}
\tilde{\Delta}(t)=\sum_{n, n^{\prime}} \exp  [ (\lambda_{n}+\lambda_{n^{\prime}}^{*}) t] |u_{R n}\rangle \langle u_{L n}|\tilde{\Delta}(0)| u_{L n^{\prime}}\rangle \langle u_{R n^{\prime}}|\,.
\end{equation}

\begin{figure}[htbp]
	\centering
	\includegraphics[width=\textwidth]{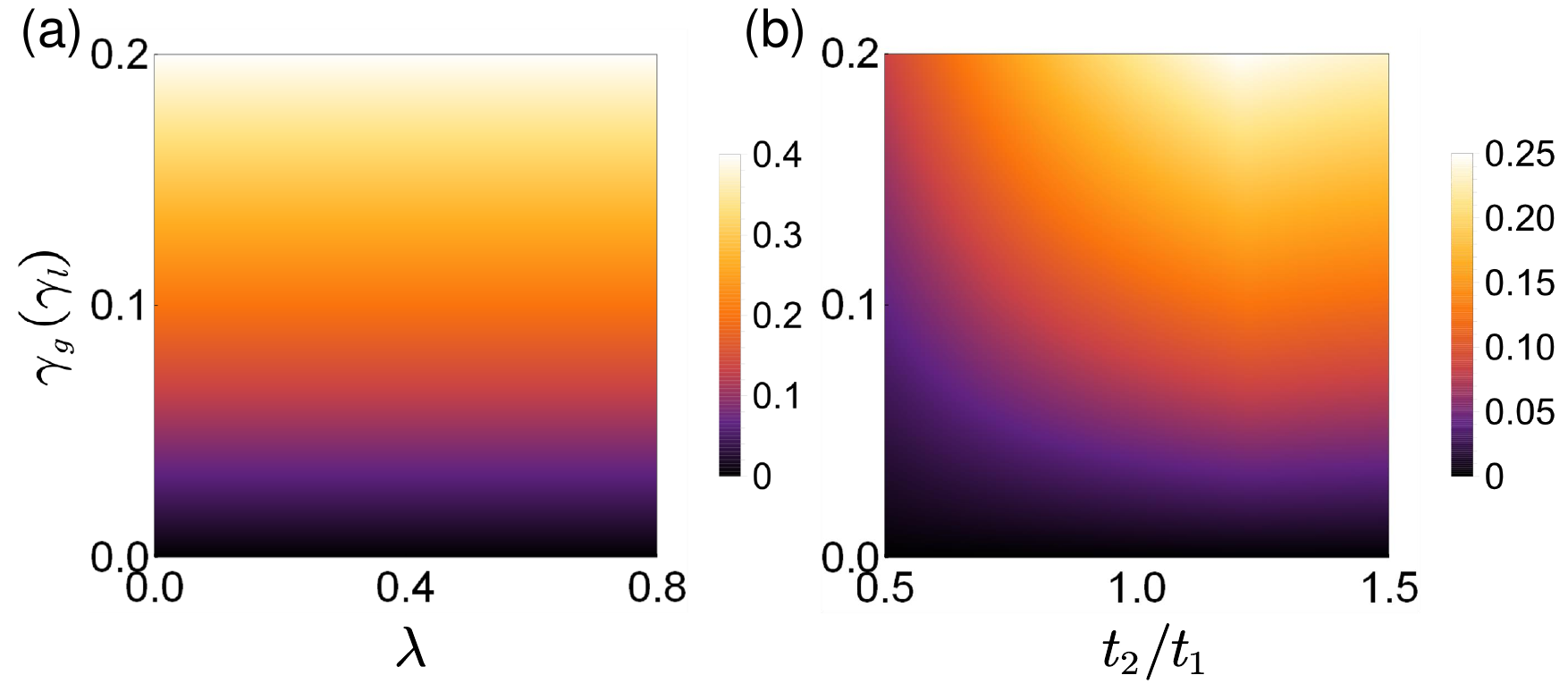}
	\caption{The phase diagrams of the the Liouvillian gap $\Lambda$ (in units of $t_1=1$) depends on the dissipation rates $\gamma_g(\gamma_l)$ and (a) the amplitude of the  intra-cell hopping modulation $\lambda$; (b) the ratio of the hopping amplitudes $t_2/t_1$. The system parameters are $\alpha=1/4$, $\delta=0$ and for (a) $t_2/t_1=1$; for (b) $\lambda=1$. Here we consider balanced loss and gain dissipation $\gamma_g=\gamma_l$. And the calculations are carried out with a finite chain with $L=60$ sites under the OBC.}
	\label{fig1}
\end{figure}

The long-time behavior of $\tilde{\Delta}$ is dominated by the sector $n=n'$. The Liouvillian spectrum $\lambda_n$ always holds negative real parts due to the dissipative nature $\mathrm{Re}(\lambda_n)\le 0$. Therefore, the long-time features of the damping dynamics are captured by the Liouvillian gap $\Lambda=\min[2\mathrm{Re}(-\lambda_n)]$. As long as the spectrum is gapped, the system fulfills exponential dissipation. Only a vanishing gap ensures algebraic convergence \cite{FSong2019}. We numerically calculate the Liouvillian gap under the open boundary condition (OBC). The results are presented in Fig. \ref{fig1}. The Liouvillian gap is dominated by the dissipation rate $\gamma$. Other system parameters, such as the amplitude of the  intracell hopping modulation $\lambda$ and  the ratio of the hopping amplitudes $t_2/t_1$ only play diminishing roles for the long-time exponential damping. However, the modulation of these parameters is essential for the transition of ways to enter the exponential stage as we will see in the next section.

\begin{figure}[htbp]
	\centering
	\includegraphics[width=\textwidth]{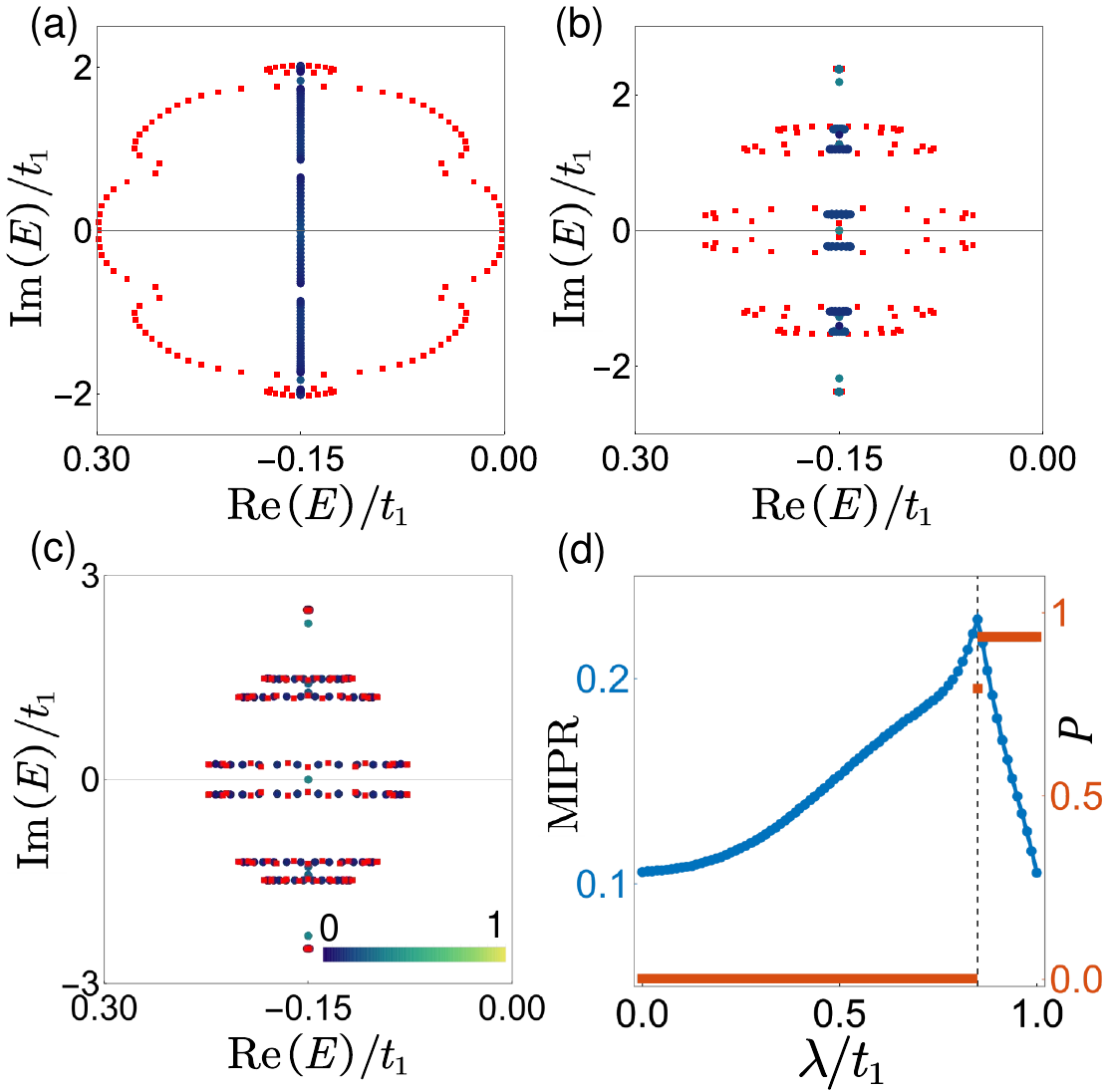}
	\caption{(a)-(c) The eigenvalues of the damping matrix on the complex plane under the PBC (red squares) and OBC (solid circles), with the color bar indicating the inverse participation ratio
(IPR) values of the corresponding eigenvectors for (a) $\lambda=0.2$, (b) $\lambda=0.85$, and  (c) $\lambda=1.0$, respectively. (d) The mean inverse participation ratio (MIPR) and the real proportion $P$ as the function of modulation amplitude $\lambda$ for OBC systems with $L=60$ sites. The dashed line indicates phase transition with the anti-$\mathcal{PT}$ symmetry breaking. The other parameters are as follows:
 $t_1=t_2=1$, $\alpha=1/4$, and $\delta=0$.}
	\label{fig2}
\end{figure}

We compare the complete Liouvillian spectra under the OBC and the PBC in Figs. \ref{fig2}(a)-\ref{fig2}(c). The discrepancy between the periodic- and the open-boundary spectra in Figs. \ref{fig2}(a) and \ref{fig2}(b) implies the failure of Bloch's theorem and the existence of NHSE, i.e., all the eigenstates of $X$ are exponentially localized at the boundary. Due to the nontrivial boundary effect from the skin modes, the conventional Fourier-transformed damping matrix $X(k)$ in Eq. (\ref{eq_xk}) does not reproduce the spectrum structure of an open chain. We use the non-Bloch theory to describe the system under the OBC via complex analytical continuation of the Bloch momentum $k \to k+i\kappa$ (or $e^{ik}\to \beta$). The non-Bloch Hamiltonian is given by,
\begin{equation}
\begin{split}
X(\beta)_{mn}=&i(\delta_{m,n-1}\overline{t}_m+\delta_{m-1,n}\overline{t}_n'+\delta_{m,1}\delta_{n,2q}t_2 \beta^{-1}\\&+\delta_{m,2q}\delta_{n,1}t_2 \beta)-\frac{\gamma}{2} \mathbb{I}_{2q}\,.
\end{split}
\end{equation}
By solving its characteristic equation $\det(\lambda \mathbb{I}_{2q}-X(\beta))=0$, we have a quadratic equation for $\beta$ with two solutions $\beta_{1,2}$ satisfying
\begin{equation}
\beta_1\beta_2=\frac{t_1't_3'\cdots t_{2q-1}'}{t_1 t_3\cdots t_{2q-1}}\,.
\end{equation}
In the continuum limit, $|\beta_1|=|\beta_2|$, which gives out $|\beta_{1,2}|=r=\sqrt{|\frac{t_1't_3'\cdots t_{2q-1}'}{t_1 t_3\cdots t_{2q-1}}|}$. Then the non-Bloch Hamiltonian can be obtained by a simple replacement of $k$ with $k+i\kappa=k+i\ln(r)$, or $e^{ik}$ with $re^{ik}$.

Notably, we find that as $\lambda$ increases, the OBC system undergoes an emergent non-Bloch anti-$\mathcal{PT}$ symmetry breaking, which is reminiscent of the non-Bloch $\mathcal{PT}$ symmetry breaking reported in Refs. \cite{Longhioe,Longhiprr}. Without loss of generality, we consider the lift of the damping matrix $X$ by a constant operation $\tilde{X}=X+\frac{\gamma}{2}\mathbb{I}$, which does not alert the spectral topology. $\tilde{X}$ possesses an emergent anti-PT symmetry, provided that $\mathcal{PT}\tilde{X}(\mathcal{PT})^{-1}=-\tilde{X}$ with $\mathcal{PT}c_{i,A(B)}(\mathcal{PT})^{-1}=c_{-i,A(B)}$ and $\mathcal{PT}i(\mathcal{PT})^{-1}=-i$. In the anti-$\mathcal{PT}$ symmetric phase, the OBC spectra of $\tilde{X}$ remain purely imaginary [see Fig. \ref{fig2}(a), but with a uniform shift on the real axis], while it becomes complex valued in the anti-$\mathcal{PT}$ symmetry broken phase [see Fig. \ref{fig2}(c) and \ref{fig2}(d)]. The phase boundary $\lambda_c=t_1-\gamma/2$ is well determined from the numerical analysis  (for more details, see Appendix \ref{appa}). To give a more complete description, we numerically calculate the real proportion for $\tilde{X}$, $P=N_r/N$, where $N_r$ and $N$ denote the number of eigenvalues with a non-zero real part and all eigenvalues, respectively. The numerical results are shown in Fig. \ref{fig2}(d). The phase transition can be transparently seen at $\lambda_c$: for $\lambda<\lambda_c$, the real proportion is almost vanishing $P=0$, but acquires large values across $\lambda_c$.

\begin{figure}[htbp]
	\centering
	\includegraphics[width=\textwidth]{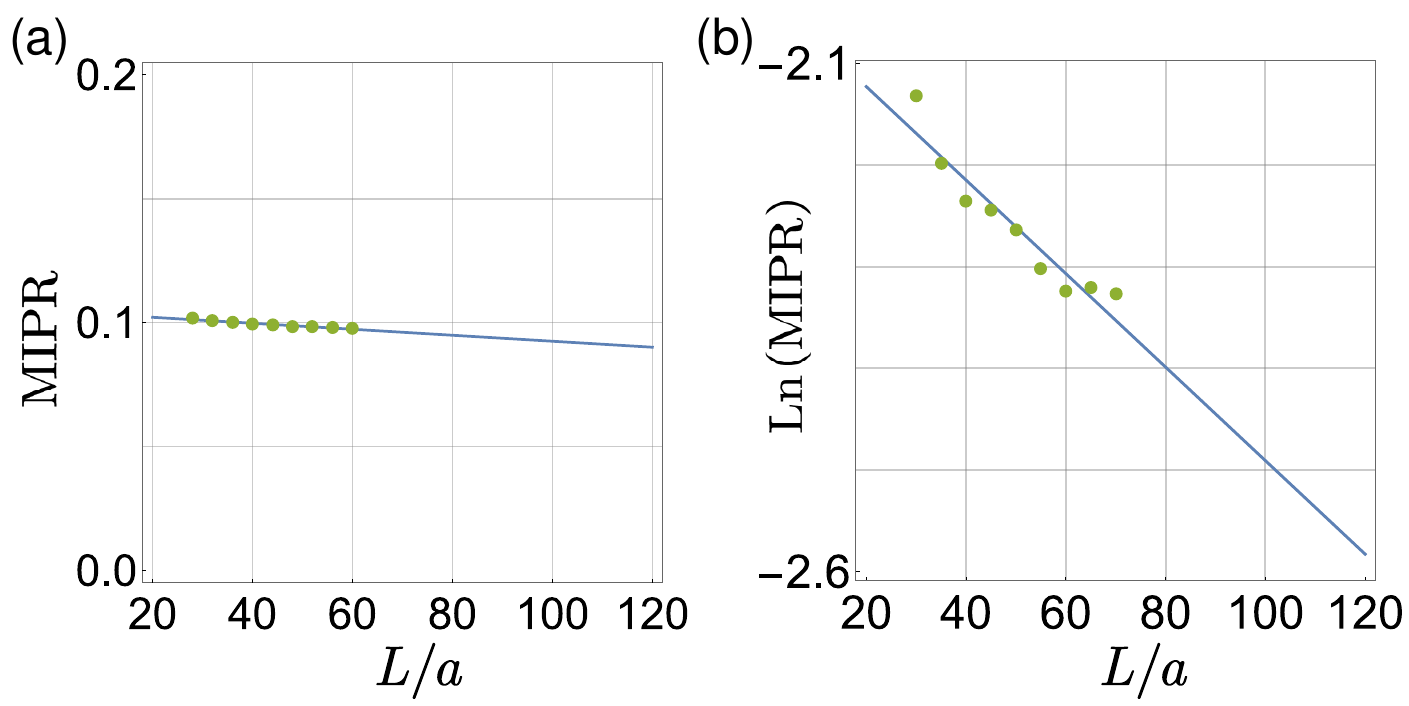}
	\caption{The MIPR as the function of system length $L$ for (a) $\lambda=0.2$ and (b) $\lambda=1$. Other parameters are the same as that in Fig. \ref{fig2}, and $a$ is the lattice distance. The dots express the numerical results for distinct system sizes, and the solid lines express the interpolation. The scale-free localization in (b) indicates immune of skin effect in the anti-$\mathcal{PT}$ symmetry broken phase.}
	\label{fig3}
\end{figure}

The phase transition signifies the change in the spectral topology of the periodic-boundary spectrum. The point gap of the spectrum (from loops to lines) closes across $\lambda_c$, then the system enters a critical phase with the states being immune to NHSE. The localization transition is demonstrated via finite-size scaling analysis, as numerically confirmed in Fig. \ref{fig3}. We calculate the MIPR for a finite open chain,
\begin{equation}
I_m\equiv \frac{1}{L}\sum_{\Lambda} I(\Lambda)\,,~I(\Lambda)=\sum_i |u_i(\Lambda)|^4/[\sum_i |u_i(\Lambda)|^2]^2\,,
\end{equation}
where $I(\Lambda)$ is the IPR for the right eigenvector $|u_R(\Lambda)\rangle$ of $X$ with eigenvalue $\Lambda$ and $u_i(\Lambda)$ being the $i$-th entry of $|u_R(\Lambda)\rangle$. As a
general rule of thumb, the IPR for an extended state is on the order of $1/L$, whereas a localized state yields finite IPR values. The MIPR behaves non-monotonically and rapidly decreases in the symmetry-broken phase as the NHSE vanishing. Typically, a skin mode $\sim e^{\kappa x}$ , where $\kappa$ is the generalized quasi-momentum, and the associated localization length $\sim 1/|\kappa|$ is independent of the system size [see Fig. \ref{fig3}(a)]. However, in the anti-$\mathcal{PT}$ symmetry-broken phase, the system exhibits quite different scaling behavior, which obeys a logarithmic rule, as shown in Fig. \ref{fig3} (b). The numerical interpolation indicates $\ln(I_m)\approx-0.22\ln(L)-1.39$. This means a scale-free localization in the symmetry-broken phase where the localization length is proportional to the system size. The finite-size scaling reveals that localization property is not dominated by the NHSE in the symmetry-broken phase. Therefore, the phase transition can be identified both by the properties of spectra and states.

We note that there are nontrivial in-gap modes in the OBC spectra [see Fig. \ref{fig2}(a)-(c)], which is characterized by a winding number (see Appendix \ref{appb}). In this paper, we focus on the bulk dynamics, thus, these topological modes do not play important roles.

\section{Damping transition}\label{sec4}
\begin{figure}[htbp]
	\centering
	\includegraphics[width=\textwidth]{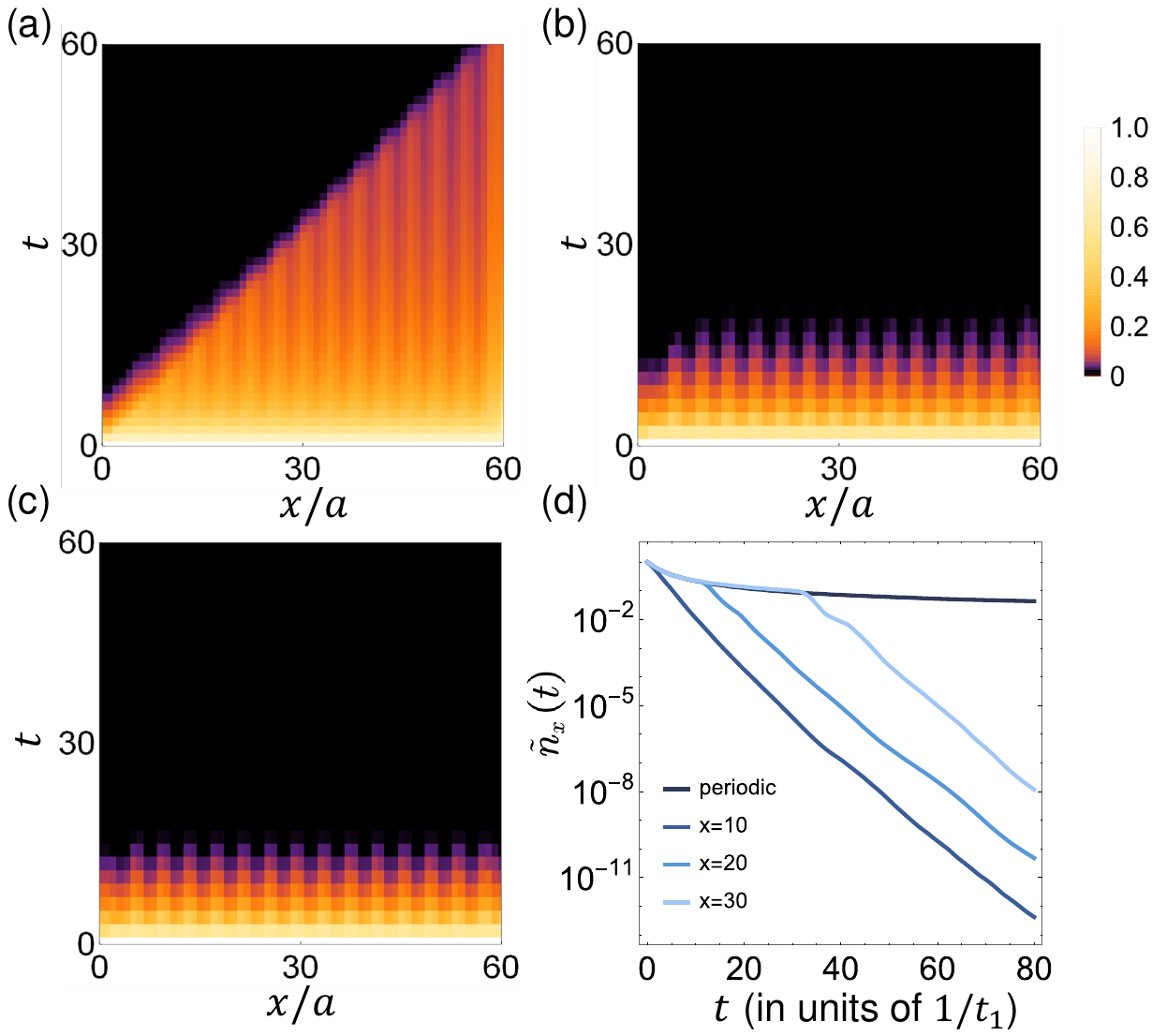}
	\caption{Time evolution of relative particle number $\tilde{n}_x(t)=n_x(t)-n_x(\infty)$ of an open chain with $L=60$ sites for (a) $\lambda=0.2$, (b) $\lambda=0.85$, and (c) $\lambda=1.0$ respectively. (d) shows sectors of (a) at certain lattice sites, compared with the results for a PBC system. Other parameters are the same as that in Fig. \ref{fig2}. The initial state is the completely filled state $\Pi_{x,s} c_{x,s}^\dagger |0\rangle$, i.e., $\Delta(0)$ is an identity matrix.}
	\label{fig4}
\end{figure}

As a manifestation of the physical implications relevant with the underlying anti-$\mathcal{PT}$ symmetry breaking, we perform full-time simulation of the damped system by numerically solving Eq. (\ref{eq_delta}). In the absence of the modulation or for weak modulation $\lambda$, the damping dynamics exhibits chiral features under the OBC, dubbed as ``chiral damping" as shown in Fig. \ref{fig4}(a). The system always enters exponential damping for a long enough evolution time due to the finite Liouvillian gap. However, the rapid exponential stage does not immediately start but follows an initial algebraic damping. The initial algebraic damping ends up successively from the left sites to the right sites, forming a sharp ``damping wave front" as more clearly seen in Fig. \ref{fig4}(d). This chiral feature fades away as $\lambda$ increases, and eventually the system only fulfills non-chiral damping across the critical point $\lambda_c$.

The chiral damping is attributed to the NHSE, according to Ref. \cite{FSong2019}. Intuitively, the asymmetric coupling effectively induces a chiral current, which is associated with the NHSE \cite{Lee2019}. To see the role of this current, we decompose the propagator $J_{x'x}=\langle xs|\exp(-iXt)|x's'\rangle$ in terms of the generalized Brillouin zone (GBZ) modes,
\begin{equation}
J_{x'x}\sim \exp(-\gamma t/2)\exp[-\kappa(x-x')]\,,
\end{equation}
where the term $\exp(-\gamma t/2)$ comes from the background damping [see the last term of  Eq. (\ref{eq_xk})], and the second term is associated with modes acquiring complex momentum $k+i\kappa$. When we can find some existent site $x'=x-\max(v_k)t(>0)$ to compensate for the background damping, we will have algebraic damping, where $v_k=\rm{Im}(\partial \lambda_\beta/\partial_k)$ is the group velocity. This indicates a wave front $x=\max(v_k)t$.

The group velocity plays important roles. However, for the modulated cases, $\lambda_\beta$ does not have a simple analytic solution. We can also estimate the group velocity by its Lebesgue measure on the imaginary axis: $v_g\sim\Delta W/2$, where $\Delta W=\sum_l \Delta W_l$ is the Lebesgue measure with $\Delta W_l$ being the bandwidh of the spectrum on the imaginary axis. As we have discussed in Sec. \ref{sec3}, the spectrum properties dramatically change across the phase transition point. When the NHSE occurs, the system possesses a continuos spectrum on the imaginary axis, which gives rise to a finite group velocity and the ballistic current. In contrast, the system possesses discrete point spectrum across $\lambda_c$ as shown in Figs. \ref{fig2}(b) and \ref{fig2}(d). The corresponding velocity is vanishing $v_g\sim\Delta W/2 \sim 0$. Intuitively, this leads to dynamical localization, thus, the damping over the lattice becomes uniform and non-chiral.

\begin{figure}[htbp]
	\centering
	\includegraphics[width=0.8\textwidth]{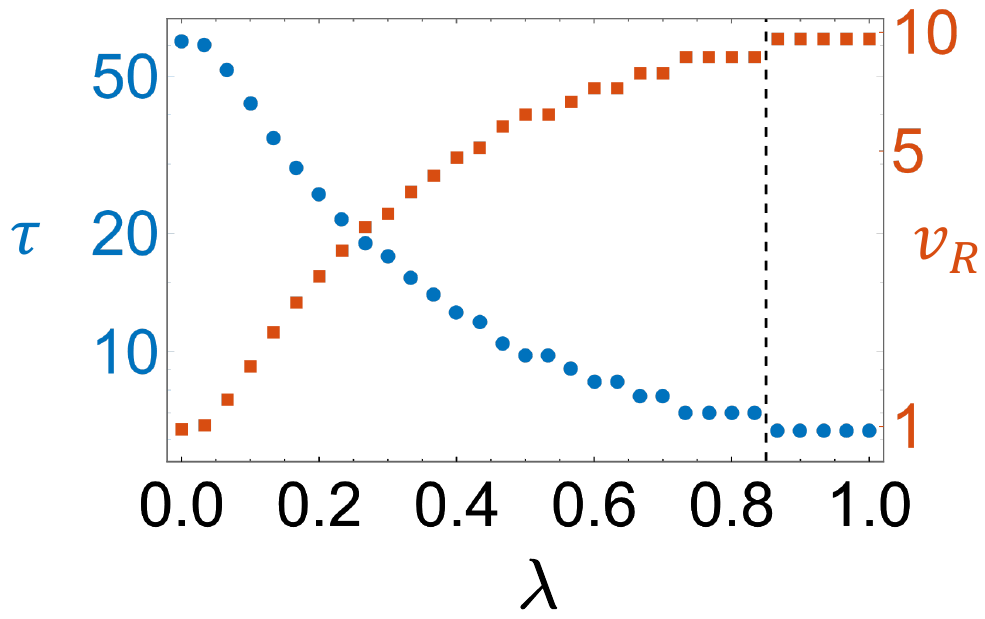}
	\caption{The longest relaxation time $\tau$ (solid circles, in unit of $1/t_1$) and relaxation velocity $v_R$ (red squares, in units of $a/t_1$) as the functions of $\lambda$. The other parameters are as follows:  $t_1=t_2=1$, $\alpha=1/4$, $\delta=0$, and $L=60$.}
	\label{fig5}
\end{figure}

To characterize the damping transition, we also calculate the longest relaxation time and relaxation velocity. The results are shown in Fig. \ref{fig5}. Equation (\ref{eq_delta}) also describes the relaxation processes of single-particle correlation. We define the relaxation time $\tau$ of the most left site as $n_{ss,L}-n_{L}(\tau)=e^{-1}n_{ss,L}$, where $n_{ss,L}$ is the density profile of the steady state \cite{Haga2021}. The relaxation velocity is defined as $v_R\equiv L/\tau$. In the non-modulated limit, the damping wavefront can be determined as $x=\max(v_k)t$ with $v_k\approx 1$ for $t_1=t_2$ \cite{FSong2019}. Therefore, it is easy to obtain that $\tau\sim L$. As $\lambda$ increases, the chiral damping fades away, and relaxation time decreases. With $\lambda$ breaking the anti-$\mathcal{PT}$ symmetry, the relaxation time $\tau \sim 1/\Lambda$. This relation between the longest relaxation time and the Liouvillian gap is also confirmed in many previous studies for some quantum dissipative system without NHSE \cite{ZCai2013,Bonnes2014,Znidaric2015}.

\section{Incommensurate case}\label{sec5}
\begin{figure*}[htbp]
	\centering
	\includegraphics[width=0.9\textwidth]{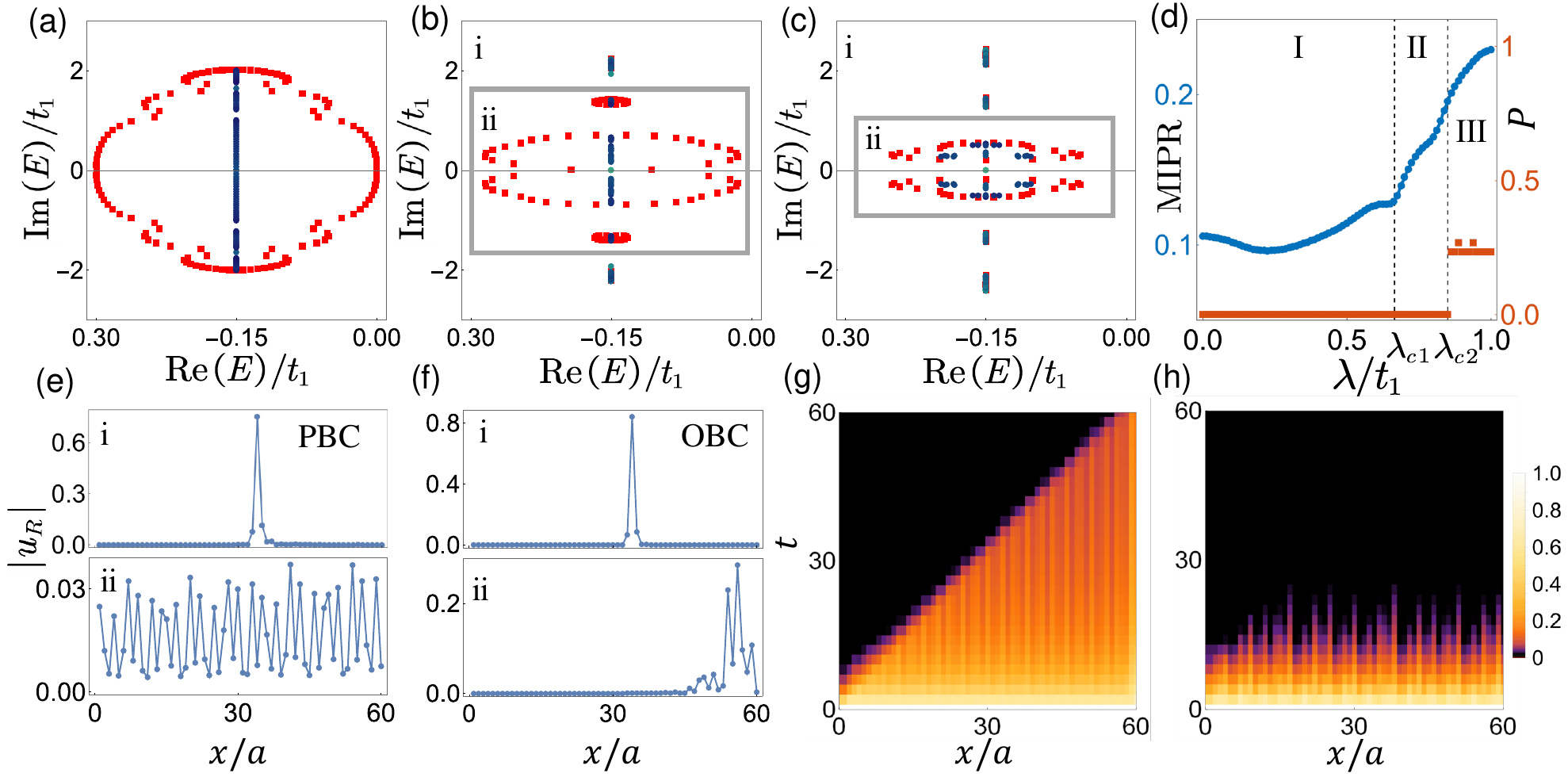}
	\caption{(a)-(c) The eigenvalues of the damping matrix on the complex plane under the PBC (red squares) and the OBC (solid circles) with the color bar indicating the IPR values of the corresponding eigenvectors for the incommensurate cases $\alpha=(\sqrt{5}-1)/2$ and (a) $\lambda=0.2$, (b) $\lambda=0.85$, and (c) $\lambda=1.0$, respectively. (d) The MIPR and the real proportion $P$ as the function of modulation amplitude $\lambda$ for OBC systems with $L=60$ sites. (e) The local density of eigenmodes for the PBC damping matrix in (b), corresponding to the eigenvalues outside (i) and inside (ii) the generalized mobility edge. (f)  The local density of eigenmodes for the OBC damping matrix in (c), corresponding to the eigenvalues outside (i) and inside (ii) the generalized mobility edge. (g) and (h) Time evolution of relative particle number $\tilde{n}_x(t)=n_x(t)-n_x(\infty)$ of an open chain with $L=60$ sites, for the incommensurate cases $\alpha=(\sqrt{5}-1)/2$, (g) $\lambda=0.2$, (h) $\lambda=1$, respectively. The gray boxes in (b) and (c) indicate a generalized mobility edge. The other parameters are as follows: $t_1=t_2=1$, and $\delta=0$.}
	\label{fig6}
\end{figure*}

We now investigate the incommensurate case. Without loss of generality, we take $\alpha=(\sqrt{5}-1)/2$ as a typical example. The quasiperiodic modulation gives rise to richer phases due to the competition between the NHSE and Anderson localization. In Fig. \ref{fig6}(d), we present the real proportion and MIPR as functions of $\lambda$, which characterize
three distinct phases, i.e., only NHSE (I, with $\lambda<\lambda_{c1}$), NHSE dominates (II, with $\lambda_{c1}<\lambda<\lambda_{c2}$), and Anderson localization dominates (III, with $\lambda>\lambda_{c2}$). The phase boundary is numerically determined, for more details, see Appendix \ref{appa}.

In phase I, the spectrum obtained under the PBC and the OBC are not identical: The spectrum under the PBC forms a loop whereas the spectrum of $\tilde{X}$ under the OBC is purely imaginary, as shown in Fig. \ref{fig6}(a). All the eigenmodes exhibit NHSE, which is similar to the commensurate case.

With the increasing $\lambda$, skin modes and localized modes coexist in phase II. The open-boundary spectrum of $\tilde{X}$ is still purely imaginary. Parts of the open-boundary spectrum are identical to the corresponding parts of the periodic-boundary spectrum, and other parts are distinct, which are separated by a generalized mobility edge. The mobility edge in Hermitian systems is defined as the energy separating localized and extended
eigenstates. For the non-Hermitian case, the spectra become complex, in general, thus, we define the generalized mobility edge as boundaries on the complex plane as shown in Fig. \ref{fig6}(b)-\ref{fig6}(c). Inside the mobility edge with open boundaries, the states are localized at the boundaries as a result of the NHSE, whereas the corresponding states under the PBC are extended due to the fact that the NHSE is sensitive to the boundary condition as presented in Fig. \ref{fig6}(e). On the contrast, outside the mobility edge, the periodic-boundary spectrum and the open-boundary spectrum are identical, implying the absence of the NHSE. The corresponding states are localized and not sensitive to the boundary condition.

In phase III, a small part of the OBC spectrum becomes complex, with corresponding states exhibiting NHSE [see Fig. \ref{fig6}(f)]. More states enter the Anderson phase outside the mobility edge,  which leads to the increase in the MIPR. When $\lambda$ is large enough, the NHSE will disappear, leaving a few extended states with complex spectra.

As presented in Fig. \ref{fig6}(g)-\ref{fig6}(h), the damping dynamics of the system with incommensurate modulation also undergoes a chiral to non-chiral transition across $\lambda_{c2}$.

\section{Realization and detection}\label{sec6}
\begin{figure}[htbp]
	\centering
	\includegraphics[width=0.9\textwidth]{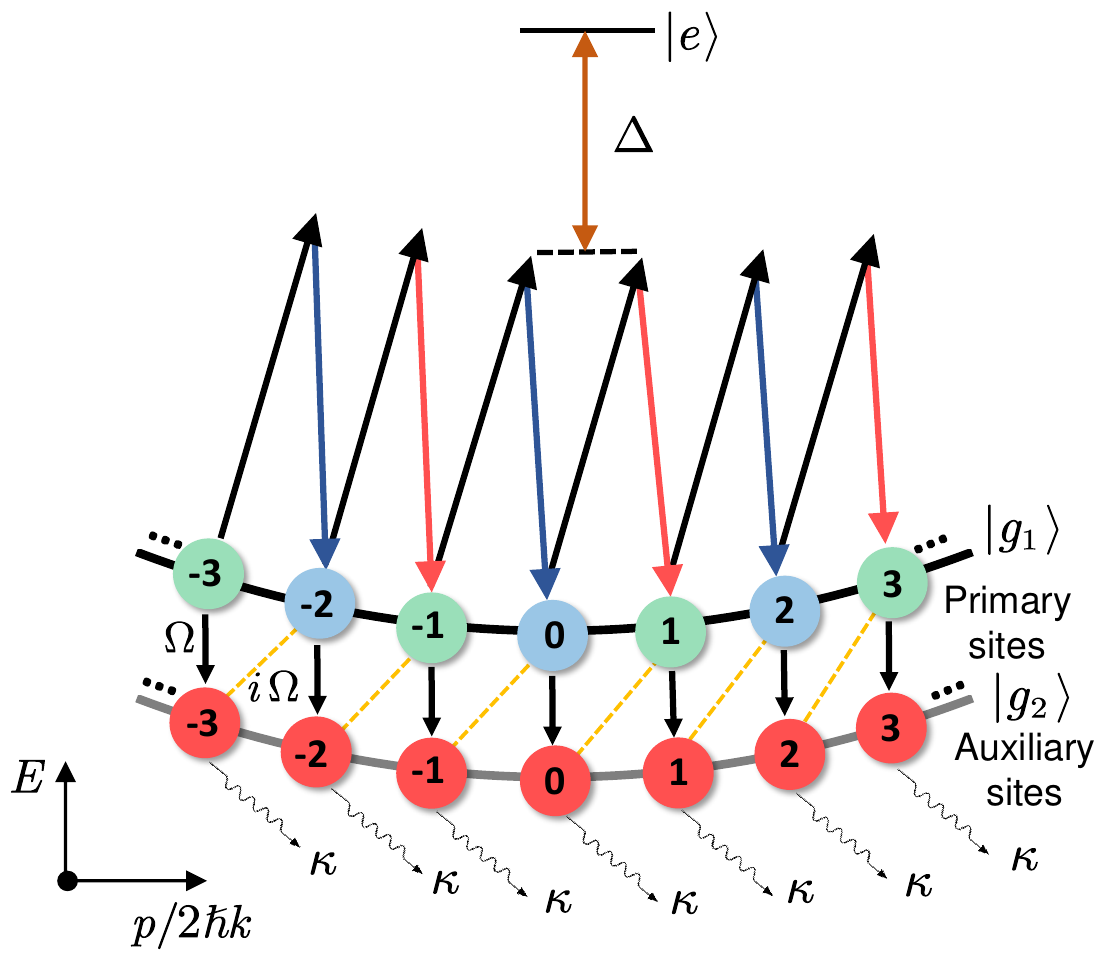}
	\caption{Schematic of our proposed experimental setup with ultracold atoms in a momentum lattice. The atoms have quadratic free-particle dispersion and are coupled by two-photon Bragg transition with resonant condition $\hbar\omega_n=(2n+1)4E_R$ with $E_R$ being the recoil energy. The Rabi coupling of laser fields $\omega_{-}^{2(n-1)}$ (illustrated by red arrows) are modulated according to the intracell hopping strength of Hamiltonian (\ref{eq_ham}). Momentum states of atoms in $|g_2\rangle$ are utilized as auxiliary sites to engineer the dissipation. The nearest-neighbor couplings between the primary sites and the auxiliary sites, indicated by dashed orange lines, can also be realized with additional Bragg transitions, which are not explicitly illustrated here.}
	\label{fig7}
\end{figure}

   We now propose a scheme to realize the generalized AAH model with ultracold atoms in a momentum lattice \cite{Gadway2015,Meier2016,Meiernc}, as illustrated in Fig. \ref{fig7}. We consider $^{87}\mathrm{Rb}$ Bose-Einstein condensate (BEC) in a crossed dipole trap. In the involved $5^2 \mathbf{S}_{1/2}$ hyperfine ground-state manifold of $^{87}\mathrm{Rb}$ atoms, momentum states in $|g_1\rangle\equiv |1,0\rangle$ encode the lattice sites while another ground state $|g_2\rangle\equiv |2,0\rangle$ with a certain decay mode is utilized as auxiliary sites to effectively engineer the desired open dynamics (here the two quantum numbers denote $F$ and $m_F$ for the levels of $^{87}\mathrm{Rb}$ atom).

Two-photon Bragg transitions are utilized to achieve the hopping terms involved in Eq. (\ref{eq_ham}). The simulated two-photon Bragg transitions are driven by pairs of counter-propagating interfering laser fields,
\begin{equation}
\mathbf{E}^{+}(\mathbf{x}, t)=\mathbf{E}^{+} \cos (\mathbf{k}^{+} \cdot \mathbf{x}-\omega^{+} t+\phi^{+}),
\end{equation}
\begin{equation}
\mathbf{E}^{-}(\mathbf{x}, t)=\sum_{n} \mathbf{E}_{n}^{-} \cos (\mathbf{k}_{n}^{-} \cdot \mathbf{x}-\omega_{n}^{-} t+\phi_{n}^{-})\,,
\end{equation}
where $\mathbf{k}^{+}=k\hat x$ and $\mathbf{k}^{-}_j=-k\hat x~\forall~j$ with $k=2\pi/\lambda$. Each pair of the Raman beams $\{\omega_{+} \oplus \omega_{-}^{n}\}$ couples momentum states $|np_0\rangle$ and $|(n+1)p_0\rangle$, where $p_0=2\hbar k$ is the total two-photon recoil momentum transferred from the light fields to the atom. Excited state $|e\rangle$ can be adiabatically eliminated from the Raman
transition for a large detuning $\delta$, then the corresponding two-photon Rabi coupling is given by
\begin{equation}
t_n=\tilde{\Omega}_{n} e^{i \tilde{\phi}_{n}}=\frac{\Omega_{n}^{*-} \Omega^{+}}{2 \Delta} e^{i (\phi^{+}-\phi_{n}^{-} )}\,.
\end{equation}
The hopping rate can be independently modulated to realize the inhomogeneous landscape in Hamiltonian Eq. (\ref{eq_ham}) by tuning the Raman coupling strength $\Omega_n^{-}$ and the phase $\phi_n^{-}$.

With coupling to dissipative auxiliary sites, we can effectively induce the gain and loss process in Eq. (\ref{eq_jump}). Each two nearest-neighbor primary sites are coupled with one auxiliary site (see Fig. \ref{fig7}). The atom loss in the auxiliary sites could be generated by applying
a radio frequency pulse to resonantly transfer the atoms in $|g_2\rangle$ to an irrelevant excited state. For a large on-site decay rate $\kappa \gg \Omega$, the  decay modes in the auxiliary lattice can be adiabatically eliminated. Thus, the effective dynamics is well described by
\begin{equation}
\dot{\rho}=-i[\tilde{H}, \rho]+\mathcal{D}[L] \rho\,,
\end{equation}
\begin{equation}
L=\sum_{\langle i,j\rangle}\sqrt{\tilde{\gamma}}(c_{i}+i c_{j})\,,\label{eq_exp}
\end{equation}
where $\tilde{\gamma}=\Omega^2/\kappa$ and $\langle i,j\rangle$ runs over all nearest-neighbor sites. We note that Eq. (\ref{eq_exp}) only contains loss dissipators, which are uniformly distributed on the lattice sites, thus, different from the staggered loss dissipators in Eq. (\ref{eq_jump}). However, the proposed experimental setup can still capture the main physics discussed in previous sections.

\begin{figure}[htbp]
	\centering
	\includegraphics[width=\textwidth]{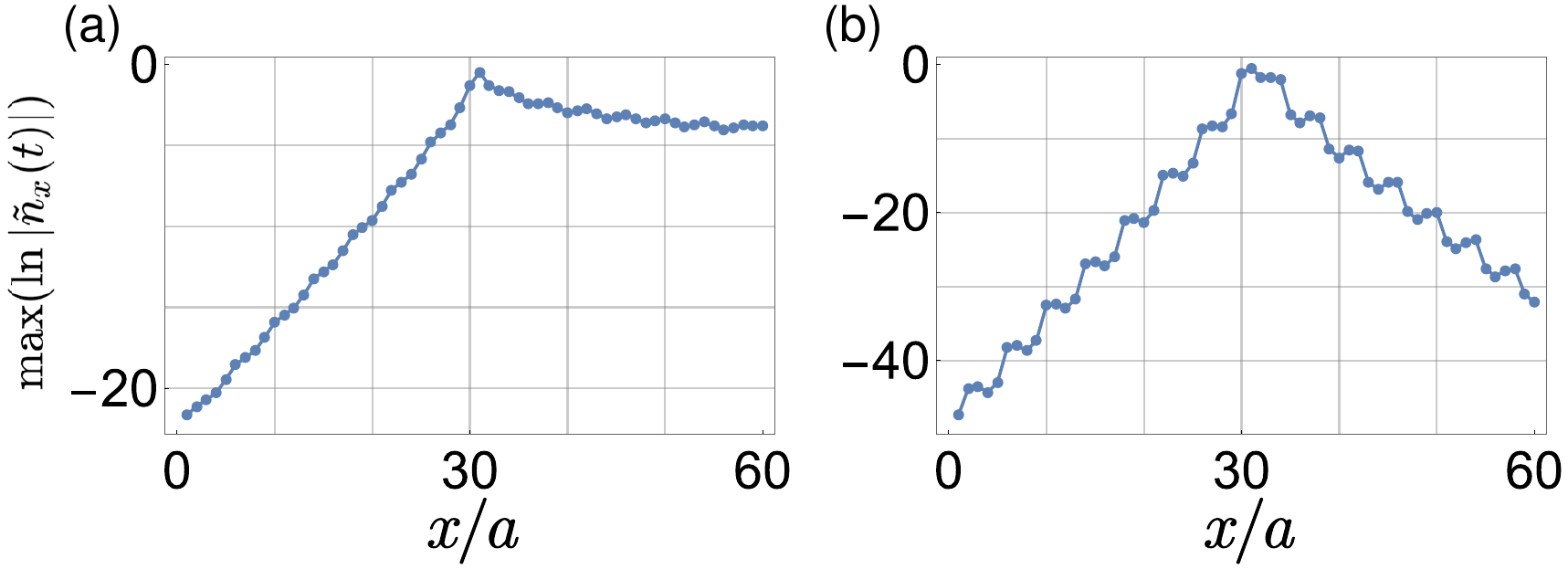}
	\caption{The maximal value of $\ln|\tilde{n}_x|$ versus lattice site $x$ in terms of the evolution time, for (a) $\lambda=0.2$ and (b) $\lambda=1.0$. Other parameters are $t_1/t_2=1$, $\tilde{\gamma}=0.15$, $\alpha=1/4$, and $\delta=0$.}
	\label{fig8}
\end{figure}

The observation of the damping dynamics revealed in Sec. \ref{sec4} requires fully filled initial state, which corresponds to a superposition of all target momentum states and is challenging to be prepared in experiments. To escape this problem, we can initialize the system in the zero momentum state, which gives a single occupation at the center site. We numerically calculate the relative particle number $\tilde{n}_x(t)$ and illustrate the maximal value at each site (in the regime of full evolution time) in Fig. \ref{fig8}. As shown in Fig. \ref{fig8}(a), the damping dynamics still exhibits chiral features for weak modulation due to the preserved NHSE. The signal strength decreases exponentially in the left segment, whereas it only fulfills a power-law decay in the right segment. This asymmetric decay rate in the opposite direction, is dubbed as the ``information constraint" \cite{CHLiuarx}. As the modulation strength increases, the phenomenon of information restrain fades away, and the signal strength becomes more symmetric in the left and right segments as shown in Fig. \ref{fig8}(b).

We can also extract the Liouvillian gap from the damping dynamics. We decompose the relative particle number in the biorthogonal eigenbasis,
\begin{equation}
\tilde{n}_x(t)=\sum_{i,j,s} e^{(\lambda_{i}+\lambda_{j}^{*}) t}\langle x, s | u_{Ri} \rangle \langle u_{Li} | u_{Rj} \rangle \langle u_{Lj} | x, s \rangle\,.
\end{equation}
For the long-time evolution, the modes with $-\mathrm{Re}(\lambda_{i}+\lambda_{j}^{*})>\Lambda_g$ can be omitted, thus we can estimate $\tilde{n}_x(t)\approx ce^{-\Lambda_gt}$. Therefore, we can extract the Liouvillian gap from the slope of $\ln(\tilde{n}_x(t))=\alpha t+\beta$ in the exponential damping stage by $\alpha \approx -\Lambda_g$. For the parameters in Fig. \ref{fig8}(a), we numerically fit the slope and obtain $\alpha\approx -1.08$, which agrees  well with the Liouvillian gap $\Lambda_g\approx 1.00$. . In experiments, the local density of state can be detected by the quasimomentum distribution $\rho(k)$ on each spin state from the time-of-flight measurement after abruptly turning off the lattice potential. The atoms in different momentum states evolve to different positions, thus the  time-of-flight measurement allows the site-resolved detection.

    We make some remarks on relevant experimental progress. The Aubry-Andr\'e model with incommensurate modulation of the on-site potential has been experimentally demonstrated with cold atoms in an optical lattice \cite{Roati2008} and coupled single-mode waveguides \cite{Kraus2012}. For the experiment with a momentum lattice, the topological Anderson insulator has been observed in disordered atomic wires\cite{Meier2018}. More recently,  a generalized Su-Schrieffer-Heeger model and a one dimensional quasi-periodic lattice have been realized \cite{DXie2019,TXiao2021}. The nonreciprocal quantum transport has been successfully observed in a dissipative momentum lattice, which makes it possible to engineer an open quantum system with atoms in a momentum lattice \cite{WGou2020}.

\section{Conclusions}\label{sec7}
To summarize, we have investigated the dynamical properties of an open generalized AAH model, in terms of the damping matrix derived from the Liouvillian superoperator. We consider both the commensurate and the incommensurate cases. When we tune strength of the modulation of hopping, the damping matrix exhibits a phase transition with anti-$\mathcal{PT}$ symmetry breaking. As the phase transition happens, the spectral topology and the localization properties dramatically change. The imaginary-to-complex transition of the spectrum only occurs when the open boundary condition is imposed, due to the NHSE. We have uncovered the physical consequence by calculating the damping dynamics of the single particle correlation function. The system undergoes a chiral to non-chiral transition as the modulation strength increases. For the incommensurate case, we identify richer phases and the existence of a generalized mobility edge. We have also proposed a possible scheme to observe these results based on the BEC in a momentum lattice. Our paper uncovers a new physical consequence of the anti-$\mathcal{PT}$ symmetry in non-Hermitian systems, and has potential implications for the experimental preparation of steady states and controlling the relaxation process without engineering the boundary conditions.

\acknowledgments
This work was supported by  the Key-Area Research and Development Program of GuangDong Province (Grant No. 2019B030330001),  the National Natural Science Foundation of China (Grants No. 12074180 and No. U1801661), and the Key Project of Science and Technology of Guangzhou (Grant No. 2019050001).

\begin{appendix}
\section{CRITICAL POINT}\label{appa}
\begin{figure}[htbp]
	\centering
	\includegraphics[width=\textwidth]{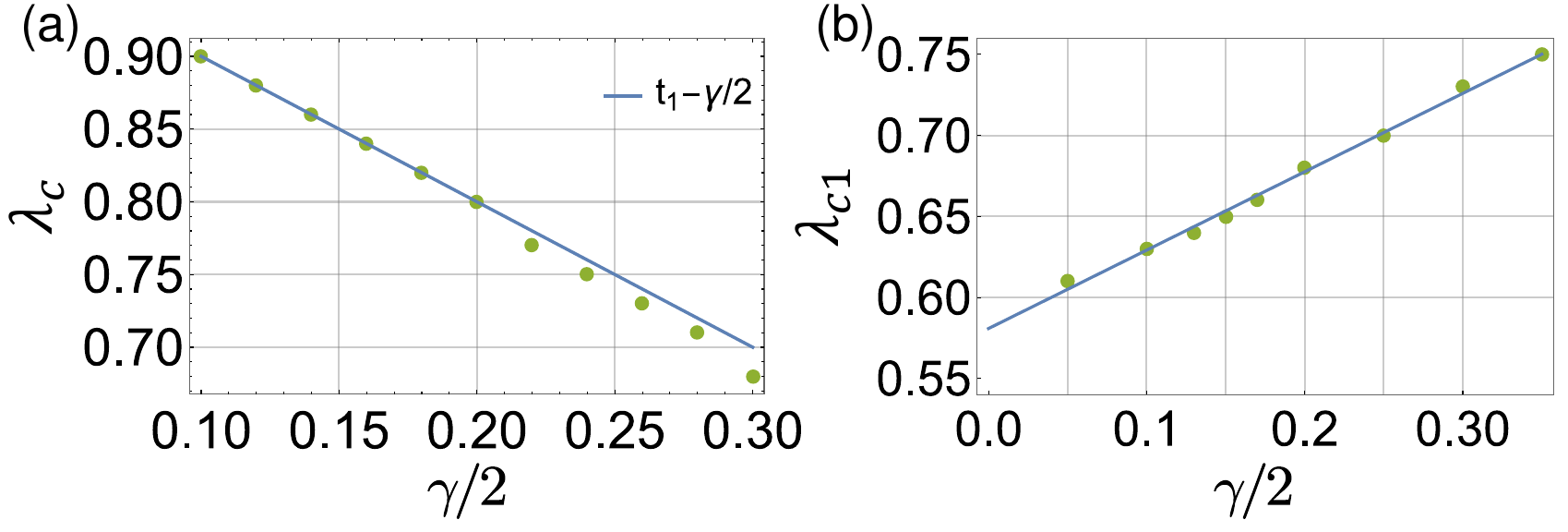}
	\caption{(a) The critical point of the anti-Grant symmetry breaking versus $\gamma/2$ for the commensurate case ($\alpha=1/4$). Here the dots show the numerical results. (b) The critical point $\lambda_{c1}$ breaking Grant $\gamma/2$ for the incommensurate case ($\alpha=(\sqrt{5}-1)/2$). The other parameters are as follows: $t_1=t_2=1$, $\delta=0$, $L=60$.}
	\label{fig9}
\end{figure}

The critical point $\lambda_c$ of the anti-$\mathcal{PT}$ symmetry breaking is numerically confirmed in Fig. \ref{fig9}(a). The critical point is extracted from the sharp change in the real proportion. After comparing different values of $\lambda_c$ by varying $\gamma$, we find a simple relation $\lambda_c=t_1-\gamma/2$ well approximately determines the phase boundary (for the case of $t_1=t_2$). And we find this relation also holds for the critical point $\lambda_{c2}$ of the incommensurate case. Furthermore, we numerically determine the critical point $\lambda_{c1}$ from the criticality of MIPR. We find that the critical point $\lambda_{c1}$ also has a simple linear dependence on $\gamma$ as shown in Fig. \ref{fig9}(b).

\section{WINDING NUMBER}\label{appb}

\begin{figure}[htbp]
	\centering
	\includegraphics[width=\textwidth]{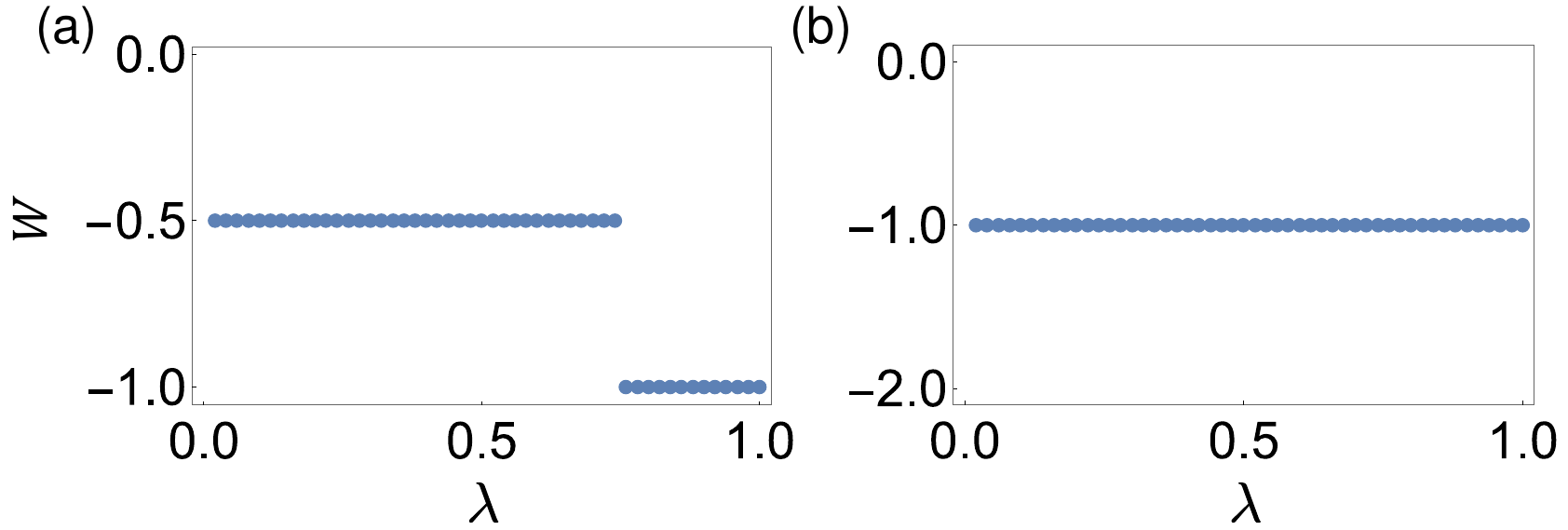}
	\caption{The winding number versus $\lambda$ for (a) $\tilde{X}(k)$ and (b) $\tilde{X}(\beta)$, respectively. Other parameters are same as that in Fig. \ref{fig2}.}
	\label{fig10}
\end{figure}

Without loss of generality, we consider the topological characterization of the compensated damping matrix $\tilde{X}$. $\tilde{X}$ has a sublattice symmetry (AIII class) $S^{-1}\tilde{X}S=-\tilde{X}$ with $S=\mathrm{diag}(1,-1,1,-1,\ldots,1,-1)$, thus, we can transform it into an off-diagonal form $\tilde{X}=[0~\tilde{X}_1;\tilde{X}_2~0]$, where
\begin{equation}
\tilde{X}_{1}\!=\!\left[\begin{array}{ccccc}
\overline{t}_{1} & 0 & 0 & \cdots & \overline{t}_{q}^{\prime} /\beta \\
\overline{t}_{2}^{\prime} & \overline{t}_{2} & 0 & \cdots & 0 \\
0 & \overline{t}_{4}^{\prime} & \overline{t}_{5} & \cdots & 0 \\
\vdots & \vdots & \vdots & \ddots & \vdots \\
0 & 0 & 0 & \cdots & \overline{t}_{q-1}
\end{array}\right]\!,~
\tilde{X}_{2}\!=\!\left[\begin{array}{ccccc}
\overline{t}_{1}^{\prime} & \overline{t}_{2} & 0 & \cdots & 0 \\
0 & \overline{t}_{3}^{\prime} & \overline{t}_{4} & \cdots & 0 \\
0 & 0 & \overline{t}_{5}^{\prime} & \cdots & 0 \\
\vdots & \vdots & \vdots & \ddots & \vdots \\
\overline{t}_{q} \beta & 0 & 0 & \cdots & \overline{t}_{q-1}^{\prime}
\end{array}\right]\!,
\end{equation}
with $\beta\in \mathcal{C}_{\mathrm{GBZ}}$; specifically, when $\beta=k$, it returns to the conventional Brillouin zone. We can define a winding number for each block \cite{Kawabata2019,QBZengprb},

\begin{equation}
w_{1,2}=\oint_{\mathcal{C}}\frac{dk}{2\pi i}\partial_k \ln \det \tilde{X}_{1,2},
\end{equation}
then we can define a wingding number for the system as $W=(w_1-w_2)/2$. We calculate the winding number for both $\tilde{X}(k)$ and $\tilde{X}(\beta)$, and the results are presented in Fig. \ref{fig10}. Due to the NHSE, conventional bulk-boundary correspondence breaks down, thus, the winding number calculated from $\tilde{X}_k$ can not correctly characterize the topology of an open chain. As shown in Fig. \ref{fig10}(b), the system always has a winding number $W=-1$, which implies that the damping transition is not topological (in terms of the band topology).

\end{appendix}

\end{document}